\documentclass[showpacs,preprintnumbers,amsmath,amssymb]{revtex4}
\usepackage{graphicx}
  \usepackage{dcolumn}
  \usepackage{bm}
   \usepackage{amsmath,amssymb}

\begin{document}
  \title{CHIRAL
  THIRRING-WESS MODEL WITH  FADDEEVIAN REGULARIZATION}
\author{Anisur Rahaman}
\email{1. anisur.rahman@saha.ac.in, 2. manisurn@gmail.com}
\affiliation{Hooghly Mohsin College, Chinsurah, Hooghly - 712101,
West Bengal, India}

 \begin{abstract}Replacing vector type of interaction of the Thirring-Wess model by the
 chiral type a new model is presented which is termed here as
 chiral Thirring-Wess model. Ambiguity parameters of
 regularization is so chosen that the model falls into the Faddeevian class.
 The resulting Faddeevian
class of model in general do not possess Lorentz invariance.
However we can exploit the arbitrariness admissible in the
ambiguity parameters to relate the  quantum mechanically generated
ambiguity parameters with the classical parameter involved in the
masslike term of the gauge field which helps to maintain physical
Lorentz invariance instead of the absence of manifestly lorentz
covariance of the model. The the phase space structure and the
theoretical spectrum of this class of model has been determined
through Dirac's method of quantization of constraint system.
\end{abstract}
  \maketitle

 \section{Introduction}
Generation of mass without violating the gauge invariance is a
  celebrated physical principle.
  In this context Schwinger model  acquired a significant position
in
 lower dimensional field theory \cite{SCH, LOW, COL, CAS, AG, APR, SADO}. Here
 photon acquires mass via a kind of dynamical symmetry breaking
  keeping the gauge symmetry of the model intact. Few years
 later,
 Thirring and Wess proposed  a two dimensional field
theoretical model where also photon acquires mass but the gauge
symmetry of the model breaks down at the classical
 level \cite{THIR}. Recently, an attempt has been made in \cite{BELV}, for systematic functional
integral bosonization of this mode. After few years of
presentation of the Thirring-Wess model, chiral
 generation of Schwinger model was proposed  in \cite{HAG}.
 However, the model remaimd less attractive over a long period because of its non-unitary problem.
 But it attracted attentions and gradually
 acquired a significant position in lower dimensional field theory after the work of Jackiw
 and Rajaraman  where they  became able to remove
 the non-unitary problem taking into account the electromagnetic anomaly into that model  \cite{JR}.
 The welcome entry of the anomaly and a suitable exploitation of the ambiguity involved therein
 made Jackiw-Rajaraman version of
 Chiral Schwinger model
 \cite{JR, ROT1, ROT2, KH, MIYAKE0, MIYAKE} along with the
 other independent regularized version of that model
 \cite{PM, MG, SUBIR0, SUBIR}
  interesting as well as attractive in lower dimensional field theory regime. Not only in the
 chiral Schwinger model but also vector Schwinger
 model \cite{SCH} turns into the so called non-confining Schwinger model
when anomaly enters into it \cite{AR}. A suitable exploitation of
the ambiguity involved here has been made to restore the lost
gauge invariance of this model in \cite{ARN}. This present work
will also be a display of exploitation of ambiguity parameter in
the so called chiral Thirring-Wess model with Faddeevian anomaly
in order to get a lorentz invariant theory where Lorentz
invariance was absent to start with.

 In the Thirring-Wess model  the authors considered  a theory of massles fermion
 interacting with massive vector
 field in two dimension.
  It can be thought of as a study of QED,
 viz., Schwinger model \cite{SCH, LOW} replacing Maxwell's field by Proca
 and that very replacement breaks the gauge symmetry at the classical
 level but a consistent field theoretical model gets birth.
 It is true that the so called non-confining Schwinger model \cite{AR, AR1}
 is a structurally equivalent gauge
 non-invariant model to the Thirring-Wess mode
 but there lies a crucial difference between these two.
 In the Thirring-Wess  mode the masslike
 term for the gauge field  was included at the classical level however
 in the the so called non-confining Schwinger model the same
 type of masslike term gets involved through one loop correction
 which contains an ambiguity parameter too. In \cite{ARN}, we have
 noticed a competition between the classically included masslike term
 and quantum mechanically generated masslike term in connection
 with gauge symmetry restoration of the so called non confining
 Schwinger model.

 An
 attempt has been made to get chiral generation of the Thirring-Wess
 mode in the similar way the chiral
 generation of the Schwinger model was made in \cite{HAG}. How anomaly becomes
 useful in the present context to get
 a consistent and physically sensible theory that we would like to
 address here for the Faddeevian class of regularization.
 So we replace vector interaction of the Thirring-Wess model by the
 chiral one that  generates a new model which we would like to term as
 chiral Thirring-Wess model.  The resulting model does not possess Physical
 lorentz invariance for
 all admissible regularization. Regularization is needed in order
 to remove the divergence of the fermionic determinant that
 appears in the process of bosonization integrating out the
 fermions one by one. It would be of interest whether the absence of physical lorentz
invariance in the
 so called Chiral Thirring-Wess model with Faddeevian class of regularization gets restored
 in a manner gauge invariance was restored
 in \cite{ARN} exploiting the arbitrariness in the ambiguity parameter.
 How does anomaly in general and regularization ambiguity in  particular come in use in this
type of investigation that is the main objective of the present
work?


\section{Chiral generation of the Thirring-Wess model with Faddeevian regularization}
 The so called Chiral Thirring Wess model can be framed by the
 following generating functional
 \begin{equation}
 Z[A] = \int d\psi d\bar\psi e^{\int d^2x{\cal L}_f}
 \end{equation}
 with
 \begin{eqnarray}
 {\cal L}_f &=& \bar\psi\gamma^\mu[i\partial_\mu + e\sqrt\pi
 A_\mu(1-\gamma_5)]\psi \nonumber \\ &=& \bar\psi_R\gamma^\mu
 i\partial_\mu\psi_R +\bar\psi_L \gamma^\mu(i\partial_\mu +
 2e\sqrt\pi A_\mu)\psi_L
  \end{eqnarray}
 Here dynamics of the $A_\mu$ field is governed by he Proca field
 and the lagrangian of which is given by
 \begin{equation}
 {\cal L}_{Praca}= \frac{1}{4} F_{\mu\nu}F^{\mu\nu} +\frac{1}{2}m^2
 A_\mu A^\mu
 \end{equation}
 Note that, we have replace the vector type of interaction $\bar\psi
 \gamma_\mu \psi A^{\mu}$ by the chiral type $\bar\psi\gamma_\mu(1+
 \gamma_{5})\psi A^{\mu}$.  Let us now proceed with the fermionic part of
 the lagrangian density. The right handed fermion remains uncoupled
 in this type of chiral interaction. So integration over this right
 handed part leads to field independent counter part which can be
 absorbed within the normalization. Integration over left handed
 fermion leads to
 \begin{eqnarray}
 Z[A] &=& \int d\psi_L d\bar\psi_L\bar\psi_L
 \gamma^\mu(i\partial_\mu + 2e\sqrt\pi
 A_\mu)\psi_L  \nonumber \\
 &=&exp\frac{ie^2}{2}\int d^2x A_\mu[M_{\mu\nu} - (\partial^\mu
 +\tilde\partial^\mu) \frac{1}{\Box} (\partial^\nu +
 \tilde\partial^\nu)]A_\nu, \label{GEN}
 \end{eqnarray}
 Where $M_{\mu\nu} = ag_{\mu\nu}$, for Jackiw-Rajaraman type of
 regularization and the model remains manifestly lorentz covariant for this setting.
 The parameter $a$ represents the
 regularization ambiguity here. In general, the elements of
 $M_{\mu\nu}$ can take any arbitrary
 values. However, the model looses both its solvability and Lorentz
 invariance in that situation. We consider here a symmetric form of
 $M_{\mu\nu}$:
\begin{equation}
 M_{\mu\nu} = \left(\begin{array}{cc} \tilde{a}  \quad \alpha \\
 \alpha  \quad  \gamma \\
 \end{array}\right)\delta(x-y).
\end{equation}
  Here regularization ambiguity is
 involved within
 the parameters $\tilde{a}$, $\alpha$ and $\gamma$. These parameters gets involved in order to remove the
 divergence in the fermionic determinant since the evaluation of the determinant needs a {\it one loop correction}
 \cite{PM, SUBIR0, SUBIR}.
 It may be the situation that all the parameters are not independent for the model to be physically sensible.

 This generating functional (\ref{GEN}) when written in terms of the auxiliary
 field $\phi(x)$ it turns out to the following
 \begin{equation}
 Z[A] = \int d\phi e^{i\int d^2x {\cal L}_B},
 \end{equation}
 with
 \begin{eqnarray}
 {\cal L}_B &=& \frac{1}{2} (\partial_\mu\phi)(\partial^\mu\phi) +
 e(g^{\mu\nu} - \epsilon^{\mu\nu)}\partial_\nu\phi A_\mu + \frac{1}{2}
 e^2 A_\mu M^{\mu\nu}A_\nu \nonumber\\
 &=& \frac{1}{2}(\dot\phi^2 - \phi'^2) + e(\dot\phi + \phi')(A_0
 - A_1) + \frac{1}{2}e^2(\tilde{a}A_0^2 + 2\alpha A_0A_1 + \gamma A_1^2).
 \end{eqnarray}
 So the total lagrangian density of our present interest is
 \begin{eqnarray}
 {\cal L} &=& {\cal L}_B + {\cal L}_{Praca} \nonumber \\
 &=& \frac{1}{2}(\partial_\mu\phi)(\partial^\mu\phi) + e(g^{\mu\nu}
 - \epsilon^{\mu\nu)}\partial_\nu\phi A_\mu + \frac{1}{2}
 e^2 A_\mu M^{\mu\nu}A_\nu  - \frac{1}{4} F_{\mu\nu}F^{\mu\nu} +\frac{1}{2}m^2
 A_\mu A^\mu \nonumber \\
 &=& \frac{1}{2}(\dot\phi^2 - \phi'^2) + e(\dot\phi + \phi')(A_0
 - A_1)
 + \frac{1}{2}e^2(\tilde{a}A_0^2 + 2\alpha A_0A_1 + \gamma A_1^2) \nonumber \\
 &+& \frac{1}{2} ({\dot A_1}^2-{A'_0}^2) +\frac{1}{2}m^2( A_0^2 - A_1^2)
 \label{PLAG}\end{eqnarray}
 Here the masslike terms for gauge fields in the lagrangian density (\ref{PLAG}) is
 \begin{eqnarray}
 {\cal L}_{mass} &=& \frac{1}{2}e^2(\tilde{a}A_0^2 + 2\alpha A_0A_1 + \gamma A_1^2)
  +\frac{1}{2}m^2( A_0^2 - A_1^2)\nonumber \\
 &=&\frac{1}{2}e^2[(\tilde{a} + \frac{m^2}{e^2})A_0^2 + 2\alpha A_0A_1 +
 (\gamma - \frac{m^2}{e^2})A_1^2].
 \end{eqnarray}
 This Lagrangian in general fails to provide Poincar\'e invariant equations
 of motion. Ambiguity in the regularization allows us to put any condition unless it
 violates
 any physical principle of the theory. We thus set $\tilde{a}+\frac{m^2}{e^2}=1$ in order to make
 the lagrangian free from the quadratic term of $A_0$. With this
 choice the constraints of the theory falls under the Faddeevian
 class \cite{FAD1, FAD2, SHATAS0, SHATAS}. It has a deeper meaning
 and interesting  consequences \cite{FAD1, FAD2, SHATAS0, SHATAS}.
 Some other choices may  lead to physically sensible theory.
 It would certainly be the issue of further investigations. Here we would like to keep ourselves
 confined in the choice that leads to Faddeevian class of constraint structure, to be
 more precise, Faddeevian class of Gauss law.

\section{Constraint analysis and determination of the theoretical spectrum}
 Let us now proceed with the constraint analysis of the theory.
 To this end we require to calculate the canonical momenta of the fields involved in the theory.
 The momentum corresponding to the field $\phi$,
 $A_0$ and $A_1$ respectively are
 \begin{equation}
 \pi_\phi = \dot\phi + e(A_0 - A_1), \label{MO1}
 \end{equation}
 \begin{equation}
 \pi_0 = 0,\label{MO2},
 \end{equation}
 \begin{equation}
 \pi_1 = \dot A_1 - A_0'.\label{MO3}
 \end{equation}
 The hamiltonian obtained through the Legendry transformation is
 \begin{equation}
 H_B = \int dx [\pi_\phi\dot\phi + \pi_1{\dot A}_1 + \pi_0{\dot
 A}_0 - {\cal L}], \end{equation} which gives the following
 hamiltonian density
 \begin{eqnarray}
 {\cal H}_B &=&  \frac{1}{2} \pi_1^2 + \pi_1A_0' + \frac{1}{2}[\pi_\phi - e(A_0 - A_1)]^2
 + \frac{1}{2}\phi'^2 - e\phi'(A_0 - A_1)\nonumber \\
  &-&  \frac{1}{2}e^2(A_0^2 + 2\alpha A_0A_1 +
 (\gamma - \frac{m^2}{e^2})A_1^2). \label{HAM0}
 \end{eqnarray}

 Equation (\ref{MO2}) is independent of $\dot A_0$. So it
 is the primary constraint of the theory. At this stage it is useful to
 work with the effective hamiltonian
 \begin{equation}
 H_{Beff}= H + \int dx u \pi_0.
 \end{equation}
 Lagrangian multiplier $u$ remains undetermined at this stage. It will be fixed later.
 The preservation of the primary constraint $\pi_0 \approx 0$, gives Gauss law as the
 secondary constraint:
 \begin{equation}
 G = \pi_1' + e(\pi_\phi + \phi') + e^2(1 + \alpha)A_1 \approx 0.
 \label{GAUS0}
 \end{equation}
 This Gauss law constraint also has to be preserve in time in order to have
 a consistent theory. The preservation condition of the Gauss law is $\dot G(x) = [G(x), H(y)] =
 0$, and that leads to the following new constraint
 \begin{equation}
 (1 + \alpha)\pi_1 + 2\alpha A_0' + (\gamma - \frac{m^2}{e^2} +
 1)A_1' \approx 0.\label{FINAL}
 \end{equation}
 The preservation of the constraint (\ref{FINAL}) does not give rise to any new
 constraint. It fixes the velocity $u$.
 Note that, the Gauss law constraint (\ref{GAUS0}) is Faddeevian in nature \cite{FAD1, FAD2, SHATAS0,
SHATAS}. Though the term 'Faddeevian'is very standard in (1+1)
dimensional field theory for the reader's benefit we should
explain a bit about Faddeevian nature of constraint. If the Gauss
constraint reflects the presence of Schwinger term like  $[G(x),
G(y)]= A \delta'(x-y)$, where $A$ is a constant, then gauge
invariance gets lost and that poses a threat on the quantization
of the theory. In Ref. \cite{FAD1, FAD2}, Faddeev initially argued
that in spite of the presence of this type of abnormality it is
possible to quantize the theory. However, the degrees of freedom
would be more in number because no gauge fixing condition is
needed. So the quantization of the present theory would be
interesting in its own right because the Gauss law G(x) gives the
following Poission bracket
\begin{equation}[G(x), G(y)]=2e^2(3 +
\alpha)\delta'(x-y)\label{POI}
\end{equation}
which fits with Faddeevian nature. In this context, we would like
to mention that if we look towards the Poisson brackets
({\ref{POI}) which would be appeared for the usual chiral
Schwinger model \cite{JR} and the
 vector Schwinger model \cite{SCH,
 AR} we will find that the Scgwinger term will be absent there. In fact, it gives a vanishing
 contribution for those cases.


 The constraints are all weak condition at this stage. If
 we impose these constraints into the hamiltonian treating them as strong condition, the hamiltonian
 will be reduced to
 \begin{equation}
 H_R = \int dx[\frac{1}{2}\pi_1^2 + \frac{1}{2e^2} \pi_1'^2 -\alpha
 \pi_1A'_1 + e(1+\alpha)A_1\phi'+\frac{1}{e}\pi_1'\phi' + \phi'^2 + \frac{1}{2}e^2[
 \alpha^2 - \gamma + \frac{m^2}{e^2}]A_1^2]. \label{RHA}
 \end{equation}
 But we have to keep it in mind that the canonical Poission brackets will be inadequate for this
 reduced Hamiltonian for computation of equations of motion \cite{DIR}. To
 get correct equations of motion for this constrained system appropriate
 Dirac brackets \cite{DIR} have to be employed in place of ordinary Poisson brackets. It
 is known that Dirac bracket
 between the two variables A(x) and B(y) is defined by
 \begin{equation}
 [A(x), B(y)]^* = [A(x), B(y)] - \int[A(x) \omega_i(\eta)]
 C^{-1}_{ij}(\eta, z)[\omega_j(z), B(y)]d\eta dz, \label{DEFD}
 \end{equation}
 where $C^{-1}_{ij}(x,y)$ is given by
 \begin{equation}
 \int C^{-1}_{ij}(x,z) [\omega_i(z), \omega_j(y)]dz = 1.
 \label{INV} \end{equation}
 Here $\omega
 _i$'s represents the second class constraints of the theory.
 With the help of equation (\ref{DEFD}), the Dirac brackets among the fields $A_1$,
 $\pi_1$, $\phi$, and $\pi_\phi$ are calculated:
 \begin{equation}
 [A_1(x), A_1(y)]^* = \frac{1}{2\alpha e^2}\delta'(x-y),
 \label{DR1}
 \end{equation}
 \begin{equation}
 [\phi(x) , \phi(y)]^* =-\frac{1}{4\alpha}\epsilon(x -y)\label{DR2}
 \end{equation}
 \begin{equation}
 [A_1(x) , \phi(y)]^* =\frac{1}{2\alpha e}\delta(x -y)\label{DR3}
 \end{equation}
 \begin{equation}
 [A_1(x) ,  \pi_1(y)]^* =\frac{\alpha-1}{2\alpha}\delta(x
 -y)\label{DR4}
 \end{equation}
 \begin{equation}
 [\pi(x) , \pi(y)]^* = -\frac{e^2}{2\alpha}(1-\alpha)^2\epsilon(x-y)\label{DR5}
 \end{equation}
 \begin{equation}
 [\phi(x) , \pi(y)]^* = -\frac{e}
 {4\alpha}(1-\alpha)\epsilon(x-y)\label{DR6}
 \end{equation}
 Making use of the equations (\ref{DR1}), (\ref{DR2}), (\ref{DR3}),
 (\ref{DR4}), (\ref{DR5}) and (\ref{DR6}) equations of motion for the
 fields appearing in the reduced hamiltonian (\ref{RHA}) are
 obtained as follows.
\begin{equation}
 \dot A_1 = \frac{\alpha-1}{2\alpha} \pi_1 - \frac{1}{2\alpha}
 (1+ \gamma -\frac{m^2}{e^2})A_1', \label{EQM1}
 \end{equation}
 \begin{equation}
 \dot \pi_1 = \pi_1' - \frac{e^2}{2\alpha}((1+\alpha)(1-\alpha^2)-
 (1 - \alpha) (\alpha^2 - \gamma + \frac{m^2}{e^2}))A_1. \label{EQM2}
 \end{equation}
 \begin{equation}
 \dot\phi = -\phi'- \frac{1}{e}\pi_1'+\frac{e}{2\alpha}
 (\gamma -2\alpha^2+1 -\frac{m^2}{e^2})
 A_1.\label{EQM3}\end{equation}
 After a little algebra, we find that the above three equation reduce to the
 following Lorentz invariant equations
 \begin{equation}
 (\Box - \frac{(\alpha-1)^2}{\alpha}e^2)\pi_1 = 0\label{SPEC0},
 \end{equation}
 and
 \begin{equation}
 \partial_+\eta = 0,\label{SPEC1}
 \end{equation}
 if we set the following relation of the ambiguity parameters $\alpha$ and $\gamma$
 with the classical parameter $m^2$
 \begin{equation}
 m^2=e^2(1+\gamma-2\alpha)\label{RES}.
 \end{equation}
The field $\eta$ in (\ref{SPEC1}) is defined as $\eta = \phi +
\frac{\alpha} {e(\alpha - 1)} (\dot A_1 + A'_1)$.

The setting of the above relation (\ref{RES}) becomes possible
without violating any physical principle if we are allowed to
exploit  the arbitrariness admissible in the
 ambiguity parameters. We are familiar with this practice in different contexts
 \cite{JR, ROT1, ROT2, KH, PM, MG, ARN,  RABIN, ABD, ARS}.
 The above settings makes the model not only solvable but also
 renders an interesting lorentz invariant theoretical spectrum,
 though to start with lorentz covariance was not manifested in the lagrangian.
 The equations
 (\ref{SPEC0}) and (\ref{SPEC1}) suggest that the theoretical
 spectrum contains a massive boson and a massless boson with a definite
 chirality.
 The square of the mass of the massive boson is given by
 $\tilde{m}^2 = -\frac{(\alpha-1)^2}{\alpha}$. The parameter $\alpha$ must be negative for
 the mass of the boson to be positive. Of course, one can set the matrix $M_{\mu\nu}$ to start with
 in such a way such that mass term comes out positive. Since the massless
 boson appeared in the spectrum has a definite chirality, it can be
 thought of as the a boson of the opposite chirality to this chiral boson has been eaten up during the
 process. The eaten up chiral boson, which is equivalent to a chiral fermion in $(1+1)$ dimension
 is, therefore, can be considered as it has became confined. This scenario would be more transparent when
 we will study this model imposing a chiral constraint in the following
 section.
\section{Imposition of chiral constrain}
Chiral boson is a basic ingredient of heterotic string theory. So
 it would be beneficial to express this model in terms of chiral
 boson. It is also a matter of investigation whether this model remains solvable after
 imposition of this constraint.
In this context, I should mention that if we impose this type of
constraint to any arbitrary model that may bring a disaster so far
Lorentz invariance and exactly solvability is concerned.
 We, therefore, impose  a chiral constraint in the model described by the
 lagrangian density (\ref{PLAG}) to express the model in terms of chiral boson in
 a manner it was done in \cite{KH} and investigation is carried out towards the  study of
 its solvability and maintenance of its Lorentz invariance.

 Let us now proceed to
 impose the following chiral constraint
 \begin{equation}
 \omega(x) = \pi_\phi(x) - \phi'(x) = 0. \end{equation} It is a
 second class constraint itself since
 \begin{equation}
 [\omega(x), \omega(y)] = -2\delta'(x-y). \end{equation}
  After
 imposing the constraint $\omega(x) = 0$, into the generating
 functional we find
 \begin{eqnarray}
 Z_{CH}&=& \int d\phi d\pi_\phi \delta(\pi_\phi - \phi')
 \sqrt{det[\omega, \omega]}e^{ i\int d^2x(\pi_\phi\dot\phi - {\cal
 H}_B)}
 \nonumber \\
 &=&\int d\phi e^{i\int d^2x{\cal L}_{CH}} ,\end{eqnarray} with
 \begin{equation}
 {\cal L}_{CH} = \dot\phi\phi' -\phi'^2 + 2e(A_0 - A_1)\phi' +
 \frac{1}{2}e^2 [(\gamma- \frac{m^2}{e^2} - 1)A_1^2 + 2(\alpha + 1)A_0A_1].
 \end{equation}
 It provides  a systematic description of the previous lagrangian (\ref{PLAG})
 in terms of chiral boson \cite{KH}.
 Note that the first two term is the kinetic term of the chiral
 boson \cite{SIG, FJ, BEL,IMB}.
 In \cite{KH}, we found the imposition of this type of chirl
 constraint
 on the usual chiral Schwinger model with one
 parameter class of regularization provided by Jackiw and
 Rajaraman and  a description of the usual chiral Schwinger model
 in terms of chiral boson resulted in. Here we have got an opportunity
 of using the same prescription once more.
 In the following section, we will carry out the
 hamiltonian analysis of the above lagrangian adding the kinetic
 energy term for the Proca field with the lagrangian density ${\cal
 L}_{CH}$. Needless to mention that the mass term for Praca field is already
 incorporated within the mass like terms of the $A$ fields. So the starting
 lagrangian density in this situation is
 \begin{equation}
 {\cal L} = {\cal L}_{CH} - \frac{1}{4} F_{\mu\nu}F^{\mu\nu}.
 \end{equation}
 Here $F_{\mu\nu}$ stands for the field strength for the
 electromagnetic field. Though this model has a structural
 similarity with the chiral Schwinger model there lies a crucial
difference between these two. Unlike the Chiral Schwinger, model
this model contains a classical parameter and we have already
seen in the previous section that that very parameter lies in the
root to make this model exactly solvable with a Lorentz invariant
theoretical spectrum getting mixed up suitability with the
ambiguity parameter.

\section{Determination of theoretical spectrum after the imposition of chiral constraint}
 For the determination of theoretical spectrum at first we need to calculate momenta corresponding
 to the field describing the theory. From the standard definition the
momenta corresponding to the field $\pi_\phi$, $\pi_0$ and $\pi_1$
are
 found out.
 \begin{equation}
 \pi_\phi = \phi',\label{SMO1}
 \end{equation}
 \begin{equation}
 \pi_0 = 0,\label{SMO2}
 \end{equation}
 \begin{equation}
 \pi_1 = \dot A_1 - A_0'.\label{SMO3}
 \end{equation}
 Using the above equations it is straightforward to obtain the
 canonical hamiltonian through a Legendry transformation which
 reads
 \begin{equation}
 H_C = \int dx[\frac{1}{2}\pi_1^2 + \pi_1A_0' + \phi'^2 - 2e(A_0 -
 A_1)\phi'  - \frac{1}{2}e^2[(\gamma- \frac{m^2}{e^2} -1)A_1^2 + 2(1 + \alpha)A_0A_1)].
 \end{equation} Equation (\ref{SMO1}) and (\ref{SMO2}) are the primary
 constraints of the theory. Therefore, the effective hamiltonian is
 given by
 \begin{equation}
 H_{EFF} = H_C + \tilde{u}\pi_0 + v(\pi_\phi - \phi'),
 \end{equation}
 where $\tilde{u}$ and $v$ are two arbitrary lagrange multiplier. The
 constraints obtained in (\ref{SMO1}) and (\ref{SMO2}) have to be
 preserve in order to have a consistent theory. The preservation of
 the constraint (\ref{SMO2}), leads to a new constraint which is the
 Gauss law of the theory:
 \begin{equation}
 G = \pi_1' + 2e\phi' + e^2(1 + \alpha)A_1 \approx 0. \label{GAUS}
 \end{equation}
 The preservation of  constraint (\ref{SMO1}), though
 does not give rise to any new constraint it fixes the velocity $v$ and that
 comes out to be
 \begin{equation}
 v = \phi' - e(A_0 - A_1). \label{VEL}
 \end{equation}
  The preservation of the Gauss law constraint $\dot G = 0$, again gives rise to a new
  constraint
  \begin{equation} (1 + \alpha)\pi_1 +
 2\alpha A_0' + (\gamma- \frac{m^2}{e^2}+ 1)A_1'\approx 0.\label{FINC}
 \end{equation}
 No new constraints comes out from the preservation of
 (\ref{FINC}). So the phase space of the theory
 contains the following four constraints.
 \begin{equation} \omega_1
 = \pi_0 \approx 0, \label{CON1}
 \end{equation}
 \begin{equation}
 \omega_2 = \pi_1' + e\phi' + e^2(1 + \alpha)A_1 \approx 0,\label{CON2}
 \end{equation}
 \begin{equation}
 \omega_3 = (1 + \alpha)\pi_1 + 2\alpha A_0' + (\gamma- \frac{m^2}{e^2} + 1)A_1' \approx
 0,\label{CON3}
 \end{equation}
 \begin{equation}
 \omega_4 = \pi_\phi - \phi' \approx 0. \label{CON4}
 \end{equation}
 The four constraints (\ref{CON1}), (\ref{CON2}), (\ref{CON3}) and
 (\ref{CON4}) are all weak condition up to this stage. Treating this
 constraints as strong condition we obtain the following reduced
 hamiltonian.
 \begin{equation}
 H_R =\int dx[ \frac{1}{2}\pi_1^2 + \frac{1}{4e^2} \pi_1'^2 + \frac{1}
 {2}(\alpha - 1) \pi_1'A_1 + \frac{1}{4}e^2[(1 - \alpha)^2 - 2(1+
 \gamma - \frac{m^2}{e^2})]A_1^2]. \label{RHAM} \end{equation}
 As has been stated earlier we need to calculate Dirac bracket in order
 to proceed for further analysis because this reduced Hamiltonian
 will give correct equations of motion only when Dirac brackets will be used
  for computation.  We find that, \\
 \noindent $C_{ij}(x,y) = [\omega_i(x), \omega_j(y)] =$
 \begin{equation}
 \left(\begin{array}{cccc}  0 & 0 & 2\alpha\delta'(x-y) & 0  \\
  0 & -2e^2(1+\alpha) \delta'(x-y) & e^2(1+\alpha)^2\delta(x-y) -\kappa\delta''(x-y)
 & e\delta'(x-y)\\
 2\alpha\delta'(x-y) & -e^2(1+\alpha)^2\delta(x-y)-\kappa \delta''(x-y) & 2(\alpha+1)\kappa \delta'(x-y) & 0\\
 0 & e\delta'(x-y) &  0  &  2e\delta'(x-y)
 \end{array}\right), \label{MAT}
 \end{equation}
 with $\kappa= 1+ \gamma - \frac{m^2}{e^2}$.
 The definition (\ref{DEFD}), along with equations
 (\ref{INV}) and (\ref{MAT}), enable us to compute the Dirac brackets
 between the fields describing the reduced Hamiltonian $H_R$:
 \begin{equation}
 [A_1(x), A_1(y)]^* = \frac{1}{2e^2}\delta'(x-y), \label{SDR1}
 \end{equation}
 \begin{equation}
 [A_1(x), \pi_1(y)]^* = \frac{(\alpha -1)}
 {2\alpha}\delta(x-y),\label{SDR2}
 \end{equation}
 \begin{equation}
 [\pi_1(x), \pi_1(y)]^* = -\frac{(1+\alpha)^2}
 {4\alpha e^2}\epsilon(x-y).\label{SDR3}
 \end{equation}
 From the reduced hamiltonian (\ref{RHAM}), the following
 first order equations of motion result in with the use of Dirac
 brackets (\ref{SDR1}), (\ref{SDR2}) and (\ref{SDR3}).
 \begin{equation}
 \dot A_1 = \frac{(\alpha-1)}{2\alpha} \pi_1 + \frac{1}{2\alpha}
 (\gamma- \frac{m^2} {e^2}+ 1)A_1', \label{SEQM1}
 \end{equation}
 \begin{equation}
 \dot \pi_1 = \pi_1' + \frac{e^2}{2\alpha}(\alpha-1)(\gamma- \frac{m^2}{e^2}+2\alpha +1)
 A_1. \label{SEQM2}
 \end{equation}
 After a little algebra, the equations (\ref{SEQM1}) and (\ref{SEQM2})
 reduce to the following
 \begin{equation}
 \partial_{+}A_1 = \frac{(\alpha-1)}{2\alpha} \pi_1 + \frac{1}
 {2\alpha}(2\alpha - \gamma + \frac{m^2}{e^2} - 1)A_1' .\label{REQ1}
 \end{equation}
 \begin{equation}
 \partial_{-}\pi_1 =\frac{e^2}{2\alpha}(\alpha-1)(\gamma- \frac{m^2}{e^2}
 + 2\alpha + 1) A_1.
 \label{REQ2} \end{equation}
Here $\partial_{\pm}$ is defined is $\partial_{\pm} =
\partial_{0} \pm \partial_{1}$.
 The above two equations (\ref{REQ1}) and (\ref{REQ2}) ultimately
 reduce to the following Klein-Gordon Equation
 \begin{equation}
 (\Box - \frac{(\alpha-1)^2}{\alpha})\pi_1 = 0\label{SSPEC},
 \end{equation}
 if we set the same relation (\ref{RES}) in the same manner
 as it was done in Sec. III.
 The equation (\ref{SSPEC}), represents a massive boson with square
 of the mass given by $\tilde{m}^2 = \frac{-(1-\alpha)^2} {\alpha }$.
  Unlike the previous situation,  no massless
 degrees of freedom appears here. Note that the constraint
 structure is different and, therefore, disappearance of massless
 degrees of freedom does not look unnatural. The results reminds us the Mitra and
 Ghosh's description \cite{MG}.  We can land on to their results
 for the specific value of the parameter $\alpha= -1$, $\gamma =
 -3$ and $m^2=0$. Here, the theoretical
 spectrum contains only a massive boson with a parameter dependent mass.
 One can think of
 it as the photon acquires parameter dependent mass and the fermions of both the
 chirality have been completely eaten up during the process.
\section{Verification of Poincar\'e algebra}
  We have already mentioned
  that the gauged lagrangian for chiral boson considered here
 does not have Lorentz covariance however it is found that the model is embedded with  a Lorentz
 invariant theoretical spectrum. So our next task is to check  Poincar\'e
 algebra  in the reduced phase space. Let us now proceed to check
 that.

 There are three elements in this algebra, the hamiltonian $H_{R}$,
 the momentum $P_{R}$ and the boost generator $M_{R}$ and they have
 to satisfy the following relation in $(1 + 1)$.
 \begin{equation}
 [P_{r}(x), H_{R}(y)]^* = 0, [M_{R}(x), P_{R}(y)]^* = -H_{R},
 [M_{R}(x), H_{R}(y)]^* = -P_{R}.\label{POIN}
 \end{equation}
Hamiltonian has already been given in (\ref{RHAM}) and the
 momentum density reads \begin{eqnarray}
 {\cal P}_R &=& \pi_1A_1' + \pi_\phi\phi', \nonumber \\
 &=& \frac{1}{4e^2}\pi_1^2 + \frac{1}{2}(1-\alpha)\pi_1A_1' +
 \frac{1}{4}e^2(1+\alpha)^2A_1^2
 \end{eqnarray} The Boost generator written in terms of hamiltonian
 density and momentum is
 \begin{equation}
 M_{R} = tP_{R} + \int dx x {\cal H}_{R} \label{MOM} \end{equation}
 Straightforward calculations shows that equation(\ref{POIN}) is
 satisfied provided the relation (\ref{RES}) between $\alpha$,
 $\gamma$ and $m^2$ is maintained.  The above calculations,
 therefore, reveals that the physical Lorentz invariance of this
 model demands the same relation (\ref{RES}) between $\alpha$,
 $\gamma$ and $m^2$.  This  certainly strengthens
the consistency of the theory under consideration. A closer
 look reveals that the Lorentz invariance is not maintained in the whole
 subspace of the theory but in the physical subspace it is
 maintained in spite of having such a deceptive appearance.

\section{Conclusion}
 In this paper we have considered the Thirring-Wess model replacing its vector
 interaction by the chiral
 one and the model resulted in, is termed here as chiral Thirring-Wess model.
 Using the standard method of quantization of constrained system by Dirac \cite{DIR},  we
 have obtained a Lorentz invariant
 theoretical spectrum provided the relation (\ref{RES}) holds. It is  fascinating
 to mention that this theory
 contains the ambiguity parameters and
the arbitrariness involved in these parameters allows us  to set
the important relation (\ref{RES}) for these parameters $\alpha$
and $\gamma$ with the classically included parameter $m^2$. We
have found that the relation (\ref{RES}),  became a crucial
ingredient to obtain the Lorentz invariant theoretical spectrum.
Thus, the physical Lorentz invariance of this theory is achieved
here by exploiting suitably the arbitrariness in the ambiguity
parameters of regularization. Note that, in \cite{ARN}, a similar
approach was made to bring back the lost gauge symmetry with the
inclusion of masslike term at the classical level.

 If we look at the theoretical spectrum we find that the photon acquires
 mass like Thirring-Wess model but the mass $\tilde{m}$ is different in this
 situation. Along with the parameter included at the classical level it also depends
  on the parameters entered  into the theory through the one loop correction.
 Fermion of a particular
 chirality gets confined here. It is not surprising since the nature of interaction and the choice
 of regularization is different in the present situation.  We found the similar situation in the
  Mitra's version
 of chiral Schwinger model, where he used a regularization that rendered a Gauss's
 law constraint from which Faddeevian type of Poission bracket resulted in \cite{PM, MG}.

 After imposing a chiral constraint into the proposed action an attempt has been made
 to obtain a new effective action  to
  describe this new model in terms of chiral boson \cite{SIG, FJ, BEL}.
It is indeed a strange at the same time an interesting aspect of
this model that after carrying out investigation on the phase
space structure we found a completely different constraint
structure from the constraint structure of the model discussed in
Sec. III,  and consequently, a drastic change in the theoretical
spectra resulted in.  Photon acquires mass as
 well  and the mass of the photon is also found to be identical to the mass
 of the massive boson as obtained in  the previous case,
 but the fermions of both the chirality are found
 to be absent, i.e., confined or eaten up during the process. If we look towards
 the structure of the theory
a deceptive appearance will be observed. To
 be precise, there  is no term in the effective action which had manifestly Lorentz
 covariant structure. However, physical Lorentz invariance is found to
 be preserved. The incredible service of the ambiguity
 parameters through the relation (\ref{RES}) has became a key concerning
 the maintenance of physical Lorentz symmetry in this case
 too. So a novel result which follows from this work is that the ambiguity in
 the regularization renders a remarkable service to make a theory Lorentz
 invariant getting mixed up suitably
with the classical parameter involved in the Proca lagrangian. If
it is asked that if the admissible arbitrariness did not permit to
set the relation (\ref{RES}) what would be the fate of this model?
Simply, it would be disaster. We will not be able to reach into
this interesting theoretical spectrum. It will not only loose its
exact solvability but loose its physical Lorentz invariance.

One more point on which we would like to emphasize is that the way
the arbitresses of the ambiguity parameter has been exploited here
to get back the Lorentz invariance of the model, has not in any,
violated any physical principle rather it has helped to maintain
the most important physical principle (Lorentz invariance) of a
physically sensible theory. Needless to mention, that the
technique is more or less standard in (1+1) dimensional QED and
Chiral QED. We have witnessed several examples of the use of this
mechanism in order to get rescued from different unfavorable as
well as un physical situations \cite{JR, ROT1, ROT2, KH, PM, MG,
RABIN, ABD, ARS}. The most remarkable one in this context is the
Jackiw and Rajaraman version of chiral Schwinger model \cite{JR},
where they saved the long suffering of the chiral generation of
the Schwinger model \cite{HAG}, from the non-unitary problem.

\noindent {\bf Acknowledgement:}
 I would like to thank the Director of
 Saha Institute of Nuclear Physics, Calcutta, and the head of the theory
 group of the same institute for providing library and computer facilities
 of the Institute.
 

\begin{thebibliography}{the}
\bibitem {SCH} J. Schwinger, Phys. Rev. {\bf 128} 2425 (1962)
\bibitem {LOW} J. Lowestein and J. A. Swieca, Ann. Phys. {\bf 68} 172 (1971)
\bibitem {COL} S. Coleman, Ann. Phys. {\bf 101} 239 (1976)
\bibitem{CAS} A. Casher, J. Kogut and L. Susskind, Phys. Rev. {\bf D10} 732 (1974).
\bibitem {AG} G. Bhattacharya, A. Ghosh, and P. Mitra, Phys. Rev. {\bf D50}  4183 (1994)
\bibitem {APR} A. Saha, A. Rahamana and P. Mukherjee, Phys. Lett. {\bf
B638} 292 (2006), Phys. Lett {\bf B643} 383 (2006)
\bibitem {SADO} F. Ardalan, M. Ghasemkhani and N. Sadoohhi, Eur. Phys. J. {\bf C71} 1606 2011
 \bibitem {THIR} W. E. Thirring and J. E. Wess,  Ann. Phys. {\bf 27} 331 1964.
 \bibitem {BELV} L. V. Belvedere and A. F. Rodrigues, Ann. Phys. {\bf 317}
 2005 423.
 \bibitem{HAG} C. R. Hagen, Ann. Phys. (N. Y.) {\bf 81} 67 (1973)
 \bibitem {JR} R. Jackiw and R. Rajaraman, Phys. Rev. Lett. {\bf 54} 1219
 1985.
 \bibitem{ROT1} H. O. Girotti, J. H. Rothe and A.K.Rothe Phys.
 Rev. {\bf 33} 514 (1986)
 \bibitem{ROT2} H. O. Girotti, J. H. Rothe and A.K.Rothe Phys.
 Rev. {\bf 34} 592 (1986)
 \bibitem {KH} K. Harada, Phys. Rev. Lett. {\bf 64} 139 (1990)
 \bibitem{MIYAKE0} S. Miyake and K. Shizuya, Phys. Rev. {\bf D46} 3781 (1987)
 \bibitem{MIYAKE} S. Miyake and K. Shizuya, Phys. Rev. {\bf D37} 2282 (1988)
\bibitem {PM} P. Mitra, Phys. Lett {\bf 284} 23 1992
 \bibitem {MG} S. Ghosh and P. Mitra, Phys. Rev. {\bf D44} 1332 1991
 \bibitem {SUBIR0} S. Mukhopadhyay and P. Mitra, Ann. Phys. (N. Y.) {\bf 241} 68 1995
 \bibitem {SUBIR} S. Mukhopadhyay and P. Mitra, Z. Phys. {\bf C67} 525 1995
 \bibitem {AR} P. Mitra and A. Rahaman, Ann. Phys. (N.Y.) {\bf 249} 34 1996.
\bibitem{ARN} A. Rahaman, Mod. Phys. Lett. {\bf A29} 1450072 2014
 \bibitem {AR1} A. Rahaman, Int. Jour. Mod. Phys. {\bf A19} 3013 2004
 \bibitem{FAD1} L. D. Faddeev, Phys. Lett. {\bf B154} 81 (1984)
 \bibitem{FAD2} L. D. Faddeev and S. L. Shatashvili, Phys. Lett. {\bf B167} 225 (1986)
 \bibitem{SHATAS0} S. L. Shatashvili, Theor. Math. Phys. {\bf 60} 770 1985, Theor. Mat. Fiz. {\bf 60} 206 (1984)
 \bibitem{SHATAS} S. L. Shatashvili, Theor. Math. Phys. {\bf 71} 336 1987, Theor. Mat. Fiz. {\bf71}  40 (1987)
\bibitem {DIR} P. A. M. Dirac, Lectures on Quantum Mechanics(Yeshiva Univ. Press New York, 1964)
\bibitem {RABIN} R. Banerjee, Phys. Rev. Lett. {\bf 56} 1889 (1986)
\bibitem {ABD} E. Abdalla, M. C. B. Abdalla, and K. D. Rorhe, 'Two Dimensional Quantum field theory' ,
(World Scientific, Singapore, 1991)
\bibitem{ARS} A. Rahaman, Phys. Lett. {\bf B697} 260 (2011)
\bibitem {SIG} W. Siegel, Nucl. Phys. {\bf 238} 307 (1984)
 \bibitem {FJ} R. Floreanini and R. Jackiw, Phys. Rev. Lett. {\bf 60} 1771 1988
 \bibitem {BEL}S. Bellucci, M. F. L. Golterman and D. N. Petcher, Nucl. Phys. {\bf B326} 307 1989
\bibitem {IMB} C.Imbimbo and J. Schwimmer, Phys. Lett. {\bf 193} 445 1987
\end{thebibliography}
 \end{document}